\documentstyle[prb,aps]{revtex} \def\narrowtext{} \tighten \twocolumn
\begin{document}
\draft

\title{On the determination of the Fermi surface in
high-$T_c$ superconductors by angle-resolved photoemission spectroscopy}
\author {
        J. Mesot,$^{1,2}$\cite{JM}
        M. Randeria,$^3$
        M. R. Norman,$^1$
        A. Kaminski,$^{2,1}$
        H.M. Fretwell,$^{2}$\cite{HMF}
        J. C. Campuzano,$^{2,1}$
        H. Ding,$^4$
        T. Takeuchi,$^{5}$
        T. Sato,$^6$ T. Yokoya,$^{6}$
        T. Takahashi,$^6$
        I. Chong,$^7$ T. Terashima,$^7$, M. Takano,$^7$
        T. Mochiku,$^{8}$
        K. Kadowaki$^{9}$
       }
\address{
         (1) Materials Sciences Division, Argonne National Laboratory,
             Argonne, IL 60439 \\
         (2) Department of Physics, University of Illinois at Chicago,
             Chicago, IL 60607\\
         (3) Tata Institute of Fundamental Research, Mumbai 400005,
             India\\
         (4) Department of Physics, Boston College, Chestnut Hill,
             MA  02467 \\
         (5) Department of Crystalline Materials Science, Nagoya
             University, Nagoya 464-01, Japan\\
         (6) Department of Physics, Tohoku University, 980-8578
             Sendai, Japan\\
         (7) Institute for Chemical Research, Kyoto University,
             Uji 611-0011, Japan\\
         (8) National Research Institute for Metals, Sengen, Tsukuba,
             Ibaraki 305, Japan\\
         (9) Institute of Materials Science, University of Tsukuba,
             Ibaraki 305, Japan\\
         }

\address{%
\begin{minipage}[t]{6.0in}
\begin{abstract}
We study the normal state electronic excitations
probed by angle resolved photoemission spectroscopy (ARPES) in
$Bi_{1.6}Pb_{0.4}Sr_{2}CuO_{6}$ (Bi2201) and
$Bi_{2}Sr_{2}CaCu_{2}O_{8+\delta}$ (Bi2212). Our main goal is
to establish explicit criteria for determining the Fermi surface
from ARPES data on strongly interacting systems
where sharply defined quasiparticles do not exist and the
dispersion is very weak in parts of the Brillouin zone.
Additional complications arise from strong matrix element variations
within the zone. We present detailed results as
a function of incident photon energy, and show
simple experimental tests to distinguish between an intensity drop due to
matrix element effects and spectral weight loss due to a Fermi crossing.
We reiterate the use of polarization selection rules in
disentangling the effect of umklapps due to the BiO superlattice in Bi2212.
We conclude that, despite all the complications, the Fermi
surface can be determined unambiguously:
it is a single large hole barrel centered about $(\pi,\pi)$ in both
materials. 
\typeout{polish abstract}
\end{abstract}
\end{minipage}}

\maketitle
\narrowtext

\section{Introduction}

The electronic structure and Fermi surfaces of conventional metals
have been studied in great detail by ARPES \cite{HUEFNER}.
The question of the determination of the Fermi surface
\cite{CAMPUZANO,OLSON,SHEN,BILAYER}
by ARPES in the normal state of the high $T_c$
superconductors is of great interest, especially since other
Fermi surface probes (like de Haas-van Alphen and positrons) have not
yet yielded useful information on the Fermi surface of the
planar Cu-O states. However, this question is not a trivial one,
since these materials are strongly correlated, and likely not
Fermi liquids, exhibiting very broad, ill-defined electronic
excitations\cite {BILAYER,KAMINSKI}.

The determination of the Fermi surface by ARPES in these systems
is further complicated by the very small dispersion in
the vicinity of the $(\pi,0)$ point of the Brillouin zone
\cite{BILAYER,GOFRON,DESSAU93}, and by
strongly ${\bf k}$-dependent photoemission
matrix elements which lead to intensity variations which have nothing
to do with Fermi crossings. Bi2212 has an additional complication:
final state diffraction of photoelectrons by the ${\bf Q}=(0.21\pi,0.21\pi)$
structural modulation in the BiO layers \cite{BILAYER,AEBI94,COMMENT}.
The combination of all these effects,
if not treated correctly, can be a source of confusion and lead to
apparently contradictory conclusions
\cite{DESSAU93,SAINI97,DESSAU99,SHEN99},
even though the data between various groups are completely consistent
with one another.

It is therefore important to establish the
criteria for unambiguously extracting the Fermi surface from ARPES data.
This is the main goal of the work presented here, in which we
study near-optimal and overdoped samples of
$Bi_{2}Sr_{2}CaCu_{2}O_{8+\delta}$ (Bi2212) and
$Bi_{1.6}Pb_{0.4}Sr_{2}CuO_{6}$ (Bi2201) using a range of
incident photon energies from 17 - 60 eV.
Our extensive study leads us to the same conclusion as our previous
work\cite{BILAYER,Bi2201}, namely that the Fermi surface consists of a
single
hole barrel centered around $(\pi,\pi)$, the most antibonding point in the
Brillouin zone. Furthermore, the Fermi surface is consistent with the
Luttinger
count with its volume scaling as one plus the number of doped holes.

The paper is organized as follows. In Section II, we
describe the samples and experimental details. We begin with ARPES
data over a wide (6eV) scale describing the full valence band in Section
III, and then turn to low energy, near $E_F$ features in the rest of the
paper. Section IV contains a brief description of polarization
selection rules and their experimental implications.
We then discuss in some detail criteria for determining the Fermi surface
in Section V. In Section VI we present data on Bi2212 and Bi2201 and show
how
the various criteria proposed in Section V fare in determining
Fermi crossings. We find that the symmetrization method for inferring
when the spectral function peak goes through the chemical potential is
a very powerful tool and works even when the energy distribution curves
(EDCs) are broad and weakly dispersive. We discuss in Section VII and VIII
the usefulness and limitations of using the integrated intensity
to determine ${\bf k}_F$, emphasizing the importance of the photon energy
and ${\bf k}$-dependence of the ARPES matrix elements. By analyzing
data obtained at different incident photon energies, we explicitly show
how one can experimentally separate matrix element effects from those
due to changes in the momentum distribution.
In Section IX, we finally turn to the Fermi surface in Bi2212
where, in addition 
to all the issues discussed above for Bi2201, one also needs to
be careful about BiO superlattice effects. Polarization selection rules
are exploited to disentangle superlattice effects from the intrinsic
CuO$_2$ electronic structure. We conclude in Section X.

An Appendix contains some further technical details related
to the symmetrization procedure.

\section{Experimental details}

Our experiments are on very high quality single crystals of Bi2212
and Bi2201 grown by the traveling solvent floating zone method with
an infrared mirror furnace, with low defect densities, as it can be
appreciated from the high resolution x-ray diffraction rocking
curve shown in Fig.~1. The samples are labeled by their doping levels
(OPT for optimal doped and OD for overdoped) together with their
onset $T_{c}$.

The as-grown Bi2212 samples are slightly overdoped and have $T_c = 87$K
with a transition width of 1K as determined by a SQUID magnetometer. These
samples are most stable in terms of their photoemission
characteristics. We look at Bi2201 samples in the doping range from
OD23K to heavily OD0K.
The samples are cleaved \textit{in situ},
and have optically flat surfaces as measured by specular laser
reflections. It is absolutely essential to characterize the flatness
of the surface on which ARPES experiments are done.
 Another measure of the sample quality, within ARPES,
is the observation of ``umklapp" bands \cite{BILAYER}
in the electronic structure of Bi2212 samples,
due to the presence of a structural superlattice
modulation in the BiO layer. Since the structural superlattice has a
periodicity of $\sim 5$ times the unit cell, very good long range order is
required for its observation.

The experiments were performed at the Synchrotron Radiation Center,
Wisconsin, using a high resolution 4-m normal incidence monochromator
with a resolving power of $10^4$ at $10^{11}$ photons/s.
The samples are carefully oriented in the sample holder to an
accuracy of 1$^\circ$ by Laue diffraction, and the orientation is
further confirmed by the observed symmetry of sharp ARPES features
around high symmetry points, as described below.

Some of the data analyzed below (in particular Figures 5, 11, 14 and
16)
were obtained using a Scienta analyzer,
at a variety of incident photon energies,
with an energy resolution of 16 meV and
a high ${\bf k}$-resolution better than 0.01 $\AA^{-1}$ at 22 eV photon
energy.
The detailed analysis presented in this paper
leads to conclusions which are completely consistent with the recent
high ${\bf k}$-resolution results of
our group \cite{HELEN} (emphasizing low temperature data on Bi2212),
as well as that of Borisenko et al. \cite{BORISENKO}.

For the Brillouin zone of Bi2212 and Bi2201, we use a square lattice
notation
with $\Gamma {\bar M}$ along the CuO bond direction, as shown in
the insets of Fig.~2. $\Gamma = (0,0)$, ${\bar M} = (\pi,0)$,
$X=(\pi,-\pi)$ and $Y=(\pi,\pi)$ in units of $1/a^*$, where $a^* =
3.83\AA$ 
is the separation between near neighbor Cu ions.
(The orthorhombic $a$ axis is along $X$ and $b$ axis along $Y$).

An example of how ARPES is used in sample alignment is shown in Fig.~2,
where spectra are shown along the
$(0,0)-(2\pi,0)$ and the $(\pi,\pi)-(\pi,-\pi)$
high symmetry lines of the Brillouin zone. This symmetry is reflected
in the position of the peak in the spectra in Fig.~2,
and allows us to accurately find the surface normal, and the angle of
the sample about this normal, completely
determining the momentum ${\bf k}$ of the outgoing electron. Note that
this 
alignment procedure only makes use of the symmetry properties of the
peak positions \cite{INTENSITY}, and does not require
a knowledge of the Fermi surface
(indicated by the curves in the top panels
of Fig.~2). 

\section{The valence band}

Our main focus will be on near $E_F$ electronic structure, but we
begin with a brief discussion of
angle-resolved photoemission from the complete valence band of Bi2212.
This covers an 
energy range of approximately 6 eV from the Fermi energy (the small peak
near zero) to the bottom of the valence band. The electronic
structure can be divided into three groups, as indicated in Fig.~3a:
the most bonding CuO$_{2}$ state is at the bottom of the valence band
(the peak at 6 eV), the antibonding state is at the Fermi energy,
and the non-bonding states are in between.
The ``lump" in the middle also includes states
from elements in the structure other than Cu and O.
But since these layers are insulating, the corresponding states do
not cross the Fermi energy.

By varying the in-plane momentum ${\bf k}$, one can map the complete
electronic structure of the valence band, as shown in Fig.~3a. These
curves were obtained without fitting peaks to the data. Instead, the
second derivative of the observed spectra, as shown in Fig.~3b, was
taken and plotted as a grey scale without any modifications
\cite{TAKAHASHI}.  
Two considerations apply: the energy step in these spectra
is only 30 meV, and therefore the details of the dispersion near
the Fermi energy are not clear, and the spectra
were obtained with a particular polarization of the photons,
so that not all states show optimal intensity. Nonetheless, the most
noteworthy features are the most bonding and antibonding states,
highlighted by thick dark lines. In the remainder of the paper,
we will focus exclusively on
the antibonding states in the region near the Fermi energy.

\section{Selection rules}

We now focus on the states crossing the Fermi energy in Fig.~3a and
show how one can determine the symmetry of the initial state in ARPES.
The ARPES intensity is governed by the (square of the) dipole matrix
element 
$M_{fi}$ connecting the initial state $\left|{\psi_i}\right\rangle$ to the
final state $\left|{\psi_f}\right\rangle$, given by
$\left|\left\langle{\psi_f}\right|{\bf A}\cdot{\bf p}
\left|{\psi_i}\right\rangle\right|^2$, where $\bf A$ is the vector
potential of the (linearly polarized) incident photon, and ${\bf p}$
is the momentum operator.
 
We use a simple version of the selection rules proposed by Hermanson
\cite{HERMANSON}. Let the photon beam be incident along a plane of
mirror symmetry of the sample (${\cal M}$).
If the detector is placed in the same mirror plane as shown in Fig.~4a,
then the final state $\psi_f$ must be even with respect to reflection
in ${\cal M}$, because if it were odd the wave function would vanish at
the detector.
The dipole transition is allowed if the entire matrix element
has an overall even symmetry. Thus two possibilities arise \cite{EXPT}.
First, if the initial state $\psi_i$
is even with respect to ${\cal M}$, then the light polarization ${\bf A}$
must also be even, i.e. parallel to ${\cal M}$. Second, if the initial
state is odd with respect to ${\cal M}$, then ${\bf A}$ must also be odd,
i.e. 
perpendicular to ${\cal M}$. This can be summarized as:
\begin{equation}
\left\langle{\psi_f}\right|{\bf A}\cdot\hat{\bf
p}\left|{\psi_i}\right\rangle
\left\{{\matrix{{\psi_i\,\,even\,\,\left\langle + \right|+\left| +
\right\rangle 
\,\,\Rightarrow \,\,{\bf A}\,\,even}\cr
{\psi_i\,\,odd\,\,\left\langle + \right|-\left| - \right\rangle
\,\,\Rightarrow \,\,{\bf A}\,\,odd}\cr
}}\right.
\end{equation}

Consider hybridized $Cu3d-O2p$ initial states, as shown in Fig.~4b,
which have a $d_{x^2 - y^2}$ symmetry about a Cu site.
These states are even with respect to $(0,0)-(\pi,0)$
(i.e. the plane defined by this symmetry axis and the z-axis)
and odd with respect to $(0,0) - (\pi,\pi)$.
Therefore, measurement along the $(0,0)-(\pi,0)$ direction will be
dipole-allowed (forbidden) if the polarization vector $\textbf{A}$ is
parallel 
(perpendicular) to this axis.  Fig.~4c shows that,
consistent with an initial state which is even about $(0,0)-(\pi,0)$,
the signal is maximized when ${\bf A}$ lies in the mirror plane
and minimized when ${\bf A}$ is perpendicular to this plane.
(The reasons for non-zero intensity in the dipole forbidden
geometry are the small, but finite, ${\bf k}$-window of the experiment
and the possibility of a small misalignment of the sample).
Similarly, we have checked experimentally that (for Bi2212 in the
Y-quadrant 
where there are no superlattice complications) the initial state is
consistent with odd symmetry about $(0,0) - (\pi,\pi)$.

While the dipole matrix elements are strongly photon energy dependent,
the selection rules are, of course, independent of photon energy.
This has been checked by measurements at 22 eV and 34 eV.
All of these results are consistent with the fact that we are probing
$Cu3d-O2p$ initial states with $d_{x^2 - y^2}$ symmetry.
In addition, as we shall emphasize below,
the selection rules can be exploited to one's great advantage
in disentangling the main CuO$_2$ ``band'' from its umklapp
images due to the superlattice in Bi2212.

\section{Fermi surface criteria}

Many criteria have been used for determining the Fermi surface
in the past without a clear discussion of
the conditions under which they are applicable.
We will present three criteria here: (A) one based on dispersion of the EDC
spectral peaks through the chemical potential, (B) a second one based on
the peak of the spectral function inferred from symmetrized data,
and (C) a third one based on rapid changes in the momentum distribution.
In the following Sections we will show how these criteria fare
when applied to experimental data.
Other criteria, not discussed in this paper, will be briefly
alluded to at the end of this Section.

The ARPES intensity is given by \cite{NK}
\begin{equation}
I({\bf k},\omega)= I_0({\bf k};\nu;{\bf A}) f(\omega)A({\bf k},\omega)
\label{intensity}
\end{equation}
for a quasi-two-dimensional system, assuming validity of
the impulse approximation. Here ${\bf k}$ is the in-plane momentum,
$\omega$ is the energy of the initial state measured relative to
the chemical potential, $f(\omega) = 1/[\exp(\omega/T) + 1]$ is
the Fermi function, and the one-particle spectral function
$A({\bf k},\omega) = (-1 / \pi)\Im mG({\bf k},\omega + i0^{+})$.
The prefactor $I_0$ is proportional
to the dipole matrix element $\left| M_{fi} \right|^2$ and thus
a function of ${\bf k}$ and of the incident photon energy $h\nu$ and
polarization ${\bf A}$.
It is also important to remember that the experimentally observed
EDC involves a convolution of the intensity of Eq.~(\ref{intensity})
with the energy resolution function and a sum over the momentum
resolution window.  There is also an additive (extrinsic) background
contribution to the EDC, however this has little effect on Fermi
surface determination since the background is negligible at the
chemical potential \cite{BACKGROUND}.
 
We first discuss the simplest case of non-interacting electrons
which have infinitely sharp energy levels leading to a spectral function
$A({\bf k},\omega) = \delta(\omega -\epsilon_{\bf k})$.
A Fermi surface crossing ${\bf k}_F$ is then defined by
the location in ${\bf k}$-space where the sharp peak of
the spectral function crosses
the chemical potential ($\omega = 0$), i.e., $\epsilon_{{\bf k}_F} = 0$.
It is also useful to look at the momentum distribution
\begin{equation}
n({\bf k})=\int_{-\infty}^{+\infty}d\omega f(\omega)A({\bf k},\omega).
\label{nofk}
\end{equation}
For non-interacting electrons $n({\bf k}) = f(\epsilon_{\bf k})$,
the Fermi function. At $T=0$ then, the momentum distribution
shows a jump discontinuity at ${\bf k}_F$. At low temperatures
there is no singularity in $n({\bf k})$ but only a rapid variation
in the vicinity of ${\bf k}_F$.

The case of interacting electrons at finite temperatures is
much more interesting. The energy levels are now broadened and
shifted by the self-energy
$\Sigma = \Sigma^{\prime}+i\Sigma^{\prime\prime}$, with
$G^{-1}({\bf k},\omega) = \omega - \epsilon_{\bf k} - \Sigma({\bf
k},\omega)$.
Thus the spectral function is given by
\begin{equation}
A({\bf k},\omega) = {1 \over \pi}
{{\left| {\Sigma^{\prime\prime}({\bf k},\omega)} \right|} \over
{\left[{\omega-\epsilon_{{\bf k}}-\Sigma^{\prime}({\bf k},\omega)}\right]^2
+ \left[{\Sigma^{\prime\prime}({\bf k},\omega})\right]^2}}.
\label{spectral}
\end{equation}
The electronic dispersion is now given by tracking the peak of the
spectral function.
We define a Fermi surface crossing by the ${\bf k}$-point at which
the spectral function peak crosses the chemical potential ($\omega = 0$)
\begin{equation}
\Re eG^{-1}({\bf k}_F,\omega=0)=0.
\end{equation}

This definition agrees with the standard definition of the Fermi surface
at $T=0$ in an interacting Fermi system which can be described by
Landau's Fermi liquid theory \cite{AGD}.
For this case, there is an additional (equivalent)
characterization of the Fermi surface in terms of a discontinuity in
$n({\bf k})$ at $T=0$.

However, it should be stressed that the discussion above in terms
of the spectral function is very general, and not necessarily
limited to Fermi liquids.
We propose to use the above definition in terms of the peak of the
spectral function, which is valid even at finite temperatures, and
use it to define the Fermi surface for high $T_c$ superconductors
even though the spectral peaks above $T_c$ are too broad for the system
to qualify as a Landau Fermi liquid \cite{KAMINSKI}.
The question of whether the $T=0$ momentum distribution shows any
singularity or not cannot be addressed experimentally, since at $T=0$
one is not in a normal state but rather a broken symmetry state.
We do find, however, that the Fermi surface we experimentally
determine above $T_c$ encloses a number of electrons which are
consistent with the Luttinger count \cite{AGD} of $(1+x)$,
where $x$ is the hole doping.

Let us now discuss in detail how these characterizations of the Fermi
surface
will be used in practice to determine ${\bf k}_F$. The first method
(A) is simply to look at the dispersion of the peaks of the measured
EDCs and determine from this when the peak position crosses the
chemical potential. There are two caveats to this method. First, the
peak of the EDC does {\it not} in general coincide with the peak of the
spectral function $A({\bf k},\omega)$. As can be seen from
Eq.~(\ref{intensity}), if there is a broad spectral function $A$ centered
about $\omega = 0$, then the peak of the EDC will be at $\omega < 0$,
produced by the Fermi function chopping off the peak of $A$, in
addition to resolution effects.
This can readily be seen in the data, as will be discussed in
connection with Fig.~5(c) below, and can be corrected for under favorable
circumstances. The second problem with method (A) is that it may
be difficult to use in cases where the dispersion is very weak, as
for instance near the $(\pi,0)$ point in the cuprates.
We should note that with sufficiently fine ${\bf k}$-sampling,
these problems are minimized, as shown recently by us in
ref.~\onlinecite{HELEN}. However, as we show here, even in the
absence of such data it is possible to make progress.

We turn to the symmetrization method (B)
which allows us to overcome both the limitations of the method (A).
This method was originally introduced by us \cite{NORMAN98}
as a means of ``dividing out the Fermi function'' from the EDC and directly
infer the spectral function $A$. In order to determine $A({\bf k}_F,\omega)$
one had to assume particle-hole symmetry on a low energy scale.
However, we now use this idea for a different purpose, namely
Fermi surface determination. As we show below we {\em do not need
any assumptions about p-h symmetry to determine} ${\bf k}_F$.
 
For an arbitrary ${\bf k}$, we define the symmetrized ARPES intensity by
\begin{equation}
I_{\rm sym}(\textbf{k},\omega)
= I(\textbf{k},\omega) + I(\textbf{k},-\omega)
\label{symm0}
\end{equation}
For simplicity, we will ignore resolution effects here,
and we use eq.~(\ref{intensity}) on the right hand side of
eq.~(\ref{symm0}).
(The effect of energy and momentum resolution
convolutions are discussed in the Appendix.)
Our goal is to use $I_{\rm sym}$ to determine
${\bf k}_F$ at which $A$ has a maximum at $\omega=0$.

Analyzing the symmetrized intensity about $\omega=0$, for any
${\bf k}$, we see that $dI_{\rm sym}/d\omega(\omega=0) = 0$ and
\begin{equation}
\left. \frac{d^2I_{\rm sym}}{d\omega^2}\right|_{\omega=0}=
\left. \frac{d^2A}{d\omega^2}\right|_{\omega=0}
-\left. \frac{1}{T}\frac{dA}{d\omega}\right|_{\omega=0}
\label{symmt}
\end{equation}
For ${\bf k} \ne {\bf k}_F$,
the second term on the right hand side dominates
at sufficiently low temperature.
For an {\em occupied} state, it is easy to see that
$dA/d\omega(\omega=0) < 0$, so that $d^2I_{\rm sym}/d\omega^2(\omega=0)$ is
positive. Thus the symmetrized intensity will exhibit a local minimum,
or a dip, at $\omega$=0 for an occupied k-state.
Conversely, for an {\em unoccupied} state $dA/d\omega(\omega=0) > 0$,
and the left hand side of eq.~(\ref{symmt}) is negative,
leading to a local maximum in the symmetrized intensity at $\omega$=0.
Precisely at ${\bf k} = {\bf k}_F$, the spectral function
has a maximum at $\omega = 0$. Thus
the second term of eq.~(\ref{symmt}) vanishes and the first term
leads to $d^2I_{\rm sym}/d\omega^2(\omega=0) < 0$ yielding a peak
or local maximum in $I_{\rm sym}$ at $\omega$=0.

In practice, symmetrization is used to determine the Fermi crossing
${\bf k}_F$ as follows.
We symmetrize {\it all} EDCs along a cut in ${\bf k}$-space and
identify ${\bf k}_F$ as the boundary in momentum space
between where symmetrized data have a dip (local minimum)
versus a peak (local maximum) at $\omega=0$.
This will be demonstrated in detail below,
where the symmetrization estimate for ${\bf k}_F$
from ARPES data is also compared with other estimates, wherever
possible, and found to agree.

Before turning to the data (in the next Section), we show one example of
a simulation which illustrates symmetrization
with resolution effects included.
In Fig.~16 we plot symmetrized intensities for five ${\bf k}$ points
along a certain cut in ${\bf k}$-space.
We see that for {\it occupied} states, corresponding to the first
two ${\bf k}$ points, the spectral functions peak at $\omega < 0$, and
thus the corresponding symmetrized spectra show dips, or local
minima, at $\omega = 0$. As ${\bf k}_F$ is approached, this minimum gets
shallower. At $k_F$, half the peak is being chopped off in the EDC.
Therefore, upon symmetrization it is completely restored.
(Note we did not build in any matrix element effects in this simulation).
Once you go beyond ${\bf k}_F$, as in the last two curves,
more than half the peak is chopped off in the EDC, so there is an
intensity drop in the symmetrized plot.
In addition, one observes that unoccupied states continue to
exhibit a peak at $\omega=0$ in their symmetrized intensity in Fig.~16,
as can be understood from the preceding discussion.

We note that symmetrization makes precise, or formalizes, a
rough criterion often used by ARPES practitioners:
at ${\bf k}_F$ a vertical
line through the chemical potential intersects the midpoint (half
the maximum intensity) of the leading edge of the EDC \cite{GAPS}.
It should be emphasized a very precise determination of the
chemical potential ($\omega = 0$) is necessary to determine
${\bf k}_F$ via symmetrization.

Finally we turn to another method (C) devised by us\cite{NK,PH}, based on
the sum rule Eq.~(\ref{nofk}) relating the energy-integrated ARPES
intensity
to the momentum distribution $n({\bf k})$.
In principle, the rapid variation of $n({\bf k})$ offers a very direct
probe of the Fermi surface, which is again not restricted to Fermi
liquids. 
(The $T=0$ momentum distribution
for known non-Fermi liquid systems, such as Luttinger liquids in
one dimension, shows an inflection at ${\bf k}_F$.)
In several cases we have demonstrated the usefulness of this method in
our earlier work \cite{NK,PH} where ${\bf k}_F$ was estimated from the
location of $\max \left| {\nabla _{\bf k}n({\bf k})} \right|$.
The same method has also been successfully used later by other authors
\cite{HUEFNER2,SCHABEL}.

However, there is an important caveat to keep in mind: one does not measure
$n({\bf k})$, but rather the integrated intensity
\begin{equation}
I(\textbf{k})=\int_{-\infty}^{+\infty}d\omega I(\textbf{k},\omega)
= I_0({\bf k};\nu;{\bf A}) n({\bf k}).
\label{integratedI}
\end{equation}
Potential problems for Fermi surface determination can, and do,
arise from the ${\bf k}$-dependence of prefactor $I_0({\bf k};\nu;{\bf A})$
due to ARPES matrix elements. In Sections VII and VIII below we
discuss in detail how to distinguish ${\bf k}$-dependences of
the integrated intensity coming from $n({\bf k})$ and from the
matrix elements.

To conclude our discussion of Fermi surface criteria, we note that we will
restrict ourselves here to the normal state. We will {\it not} discuss in
this 
paper
the notion of the ``minimum gap locus'' \cite{PH,UDFS} in a gapped state
(either below $T_c$ or in the pseudogap regime) which is a measurement of
the underlying Fermi surface that got gapped out.
We also mention, for completeness,
two other methods of Fermi surface determination, which we
will not discuss in this paper. The first one exploits the
approximate sum rule \cite{NK} that $dn({\bf k}_F)/dT = 0$, i.e.,
the integrated intensity at ${\bf k}_F$ is independent of temperature.
The second method uses the constant $\omega$ scan, or momentum
distribution curve (MDC), as a function of ${\bf k}$, at $\omega = 0$;
see ref.~\onlinecite{KAMINSKI2}.

\section{Fermi Crossing from symmetrization}

We will show below that the symmetrization method provides a simple
and general way of determining a Fermi crossing, even when the
dispersion is very small. As discussed above, we will
identify ${\bf k}_F$ as that ${\bf k}$ for
which the symmetrized data {\em first} shows a clear peak at the chemical
potential 
($\omega = 0$).

We begin with the simplest Fermi crossing along
the zone diagonal, $(0,0) \to (\pi,\pi)$, where the electronic
structure shows rapid dispersion. The data for a Bi2212-OD88K sample
shown in Fig.~5(a) were obtained with an incident
photon energy of 22 eV and polarization ${\bf A}$
parallel to the $(0,0)-(\pi,-\pi)$ axis. These $T= 180$K spectra are
extremely broad, and not at all consistent with a Fermi liquid
picture \cite{KAMINSKI,KAMINSKI2}.
We first determine the Fermi crossing ${\bf k}_F$ using the
dispersion of the observed peak.
In Fig.~5(c) we plot the observed peak positions of the
data in (a). One can see that in
the narrow ${\bf k}$ interval plotted in Fig.~5(c), the dispersion is
linear over much of the range, deviating from linearity due to
the effect of the Fermi function (as discussed above).
The extrapolation of the linear part crosses the chemical potential at
$\textbf{k}=(0.375\pi,0.375\pi)$, which is then the estimated ${\bf k}_F$
from the dispersion.

In Fig.~5(b) we plot the symmetrized intensities
$I_{\rm sym}(\textbf{k},\omega)$ obtained from the data of Fig.~5(a). From
this we see that the inferred spectral function
is peaked at $\omega=0$ also at $\textbf{k}=(0.37\pi,0.37\pi)$.
We thus find that in this case the dispersion and symmetrization
methods give identical ${\bf k}_F$ estimates.

In the following we will first look at the simpler case of Bi2201.
In the Pb-doped Bi2201 compounds there are no observable
complications arising from umklapp bands as in oxygen-doped Bi2201 or
Bi2212 (which is discussed in more detail in Section IX).
We now move along the Fermi surface in Fig.~6, where the plotted data
were obtained for an OD23K sample with an incident
photon energy of 22 eV and polarization ${\bf A}$
parallel to the $(0,0)-(\pi,0)$ axis. One can
observe a clear trend: as the $(\pi,0)-(\pi,\pi)$ line is approached,
the dispersion becomes very small. In the vicinity of the
$(\pi,0)$ it becomes very difficult to use the dispersion criterion
to determine ${\bf k}_F$. Nevertheless, symmetrized data provide
completely unambiguous results as can be seen from Fig.~7. From the
top panels (a,b, and c) of Fig.~7 one can determine precisely the Fermi
crossing
from the ${\bf k}$-point at which the spectral function inferred from
symmetrized data first peaks at the chemical potential.

Two points should be noted about the ${\bf k}_F$ estimate from
symmetrization. First, just before approaching ${\bf k}_F$ from the occupied
side,
we expect that the symmetrized data will show two peaks and a small dip,
which would be broadened by resolution into a rather flat topped
symmetrized  spectrum.
Second, there should be an intensity drop upon crossing ${\bf k}_F$
in the symmetrized spectrum, assuming that the matrix elements are not
strong functions of ${\bf k}$. Both of these effects are clearly
seen in the data.

It is equally important to be able to determine the {\it absence} of a
Fermi crossing in a cut in ${\bf k}$-space as shown in Fig.~7(d,e).
In this respect, the raw data along $(0,0)-(\pi,0)-(2\pi,0)$ is quite
difficult to interpret, since the ``band'' flattens while approaching
$(\pi,0)$, and remains extremely close to $E_F$.
Nevertheless, it is simple to see that at no point along this cut do
the symmetrized data show a peak centered at $E_{f}$, thus establishing
the absence of a Fermi crossing along this cut.

The important conclusion from this discussion is that for
Pb-doped Bi2201 one can
continuously follow a Fermi surface contour which traces a hole barrel
centered at $(\pi,\pi)$. Analysis of a large set of data using the
symmetrization method allows an unambiguous determination of the Fermi
surface crossing with high accuracy, even in the unfavorable case of
broad peaks with weak dispersion.
 
\section{Matrix elements}

As discussed at the end of Section V, great care must be used
in determining a Fermi crossing from
the integrated intensity which is the momentum distribution $n({\bf k})$
multiplied by the prefactor $I_0({\bf k};\nu;{\bf A})$. A
loss of integrated intensity as a function of ${\bf k}$ can arise
either from a drop in $n({\bf k})$ related to a Fermi crossing, or
from the ${\bf k}$-dependence of the matrix elements in $I_0$.

One possibility is to have {\it a priori} information about the
matrix elements from electronic structure calculations \cite{BANSIL}.
But as we now show, even in the absence of such information, one
can experimentally separate the effects of a strong ${\bf k}$-variation
of the matrix element from a true Fermi surface crossing.
The basic idea is to exploit the fact that changing the incident
photon energy one only changes the ARPES matrix elements and not
the momentum distribution of the initial states.

In Fig.~8(a) we shows cuts along $(0,0)-(\pi,0)$ and $(\pi,0)-(\pi,\pi)$
obtained at a photon energy of 34 eV (to be contrasted with the data in
the previous two figures for the same sample at 22 eV). The symmetrized data
are shown in
Figs.~8(b) and (c) from which we see results entirely consistent with those
obtained at 22 eV. There is no Fermi crossing along $(0,0)-(\pi,0)$
since the symmetrized data in Fig.~8(b) never show a peak at $\omega=0$.
Turning to the $(\pi,0)-(\pi,\pi)$ direction, the symmetrized
data (Fig.~8(c)) do show a Fermi crossing occurring at k=0.12$\pi$,
in agreement with the data obtained at 22 eV (k=0.14$\pi$).

However, there is an important difference between the
data sets at 22 eV and 34 eV photon energies, which can be
appreciated in Fig.~9, where the integrated intensity along the two
directions is displayed.
While at 22 eV the maximum intensity occurs
close to $(\pi,0)$ and decreases both towards $(0,0)$ and $(\pi,\pi)$,
the data taken at 34 eV show a strong depression of intensity on
approaching $(\pi,0)$, resulting in a shift of the intensity maximum
away from $(\pi,0)$. This loss of intensity cannot be interpreted
as a Fermi crossing, since at no point in the symmetrized data from
$(0,0) \to (\pi,0)$  (Fig.~8) is there a peak centered at $\omega=0$.
(In fact, this loss of intensity close to $(\pi,0)$ is responsible
for the reduced signal-to-noise ratio in the 34 eV data in Fig.~8
compared with the 22 eV data in Figs.~6 and 7).

We would like to attribute this loss in intensity around $(\pi,0)$
at 34 eV, and in fact the entire variation seen in Fig.~9(a), to
strong ${\bf k}$-dependent matrix element effects.
A direct proof is found from the data:
the EDCs at the same point in the Brillouin zone obtained at
the two different photon energies exhibit exactly the
same lineshape, i.e. one can be rescaled onto the other
as shown in Fig.~9b.

We emphasize that the results of Fig.~9 imply that the photon
energy dependence of the ARPES data is {\it not} a $k_z$ dispersion
effect. If this were the case, different incident photon energies would
be probing initial states with different $k_z$ values.
However, the scaling of Fig.~9(b) proves that it is the {\it same}
two dimensional ($k_z$-independent) initial state which is being probed,
and the photon energy dependence arises from the different
final states that the matrix element couples to.

To further illustrate the role of matrix elements, we present a model
calculation and compare it
with the data of Fig.~9. The purpose of this exercise is to determine 
whether or not the data are consistent with a matrix element variation with
${\bf k}$.
In Fig.~10(a) we use dotted lines to
show the dispersion of a model ARPES spectrum along $(0,0)$
to $(\pi,0)$ where the ``band'' approaches $E_F$ near $(\pi,0)$
without a Fermi crossing.
(The dispersion is chosen from the tight binding fit
of Ref.~\onlinecite{NORM95} to ARPES data on Bi2212 for illustrative
purposes
even though we will compare it to data on Bi2201).
The full curves in Fig.~10(a) show the effect of the matrix element
variation on the model spectra, simulated by $I_0({\bf k}) =
\sin^{4}(0.6k_{x})$ along $(0,0)$ to $(\pi,0)$.
This is a simple phenomenological matrix element (squared)
which satisfies the following properties:
it vanishes at the $\Gamma$ point, as dictated by
symmetry, and then has non-monotonic behavior along
$\Gamma$-M with a peak away from the M point, similar
to that obtained by detailed band-structure
calculations \cite{BANSIL}. We emphasize that,
beyond this, no deep meaning should be attached to
the simple analytical form used.

Fig.~10(b) shows the momentum dependence of the
intensity integrated over the (large) energy range of -350 to +50
meV as a solid line, and over the (narrow) energy range -50 to +50 meV as
a dashed line.  One can see that this simple example shows remarkable
agreement with the measured intensity using the same integration
ranges, shown in Fig.~10(c). These results are easy to understand. The
data with the large integration range first increase in intensity
simply following the matrix element variation. The data with
integration over a small energy range however shows a different
intensity behavior. They start to increase for k-values higher
than 0.65$\pi$ because only when the peak is closer to the Fermi energy
does it contribute to the integrated intensity. It then again
decreases rapidly because of the strong decrease of the matrix
element, as in the previous case.

However, not recognizing the role of
matrix elements, some authors have ascribed the differences between
the 22 eV and 34 eV photon energy data to
additional ``mysterious'' states around $(\pi,0)$ \cite{DESSAU99}.
We feel that there is no necessity to invoke such states or
to assert changes in Fermi surface topology with photon energy.
In conclusion, all our data on Bi2201 when analyzed using the methods
described above indicate a large Fermi surface centered around $(\pi,\pi)$
\textit{independent of the photon energy}.

\section{The Momentum Distribution}

Despite the matrix element issues discussed above, it is
nevertheless interesting to study the integrated intensity
(Eq.~(\ref{integratedI})) for a dense data set in the entire zone.
The results obtained are shown in Fig.~11 for an OD0K sample.
In the top panels Figs.~11(a) and (c) we show
the integrated intensity $I({\bf k})$ around the $(\pi,0)$ point
obtained at two different photon energies: 22 eV and 28 eV respectively.
In the lower panels Figs~11(b) and (d), we plot the magnitude of
the logarithmic gradient: $|\nabla_{\bf k}I({\bf k})|/I({\bf k})$
which emphasizes the rapid changes in the integrated intensity.
The logarithmic gradient filters out the less abrupt changes in
the matrix elements and helps to focus on the intrinsic variations
in $n({\bf k})$. This can be seen from the fact that the integrated
intensities in the top panels are quite different for 22 eV and 28 eV,
while the logarithmic gradients are much more similar.

The Fermi surface can be clearly seen as two high intensity arcs
curving away from the $(\pi,0)$ point. Modulo matrix element effects,
the results obtained at the two different photon energies are
quite similar. Moreover, the Fermi surface estimated by this method
is in good agreement with the one obtained from the symmetrization
analysis above.

\section{The Fermi surface of B\lowercase{i}2212}

We now turn to Bi2212 where, in addition to all the issues
discussed above for Bi2201, there is
an added complication due to the presence of
umklapp bands arising from the superlattice modulation
with wavevector ${\bf Q} = (0.21\pi,0.21\pi)$ in the BiO layers.
Another difference with Bi2201, which we will not address here, is that
Bi2212 has a CuO$_2$ {\it bilayer}; we only mention that no bilayer
splitting is observed in the ARPES data on Bi2212 as discussed in detail
in Ref.~\onlinecite{BILAYER}.
In this Section we will first review the effect of the superlattice
on the electronic structure probed by ARPES, emphasizing the usefulness of
polarization selection rules.
In an earlier letter \cite{BILAYER}, we had shown data along the
principal axes of Bi2212. Here we present cuts
throughout the Brillouin zone, analyzed using the symmetrization and
integrated intensity methods discussed above, together with a detailed
study of the photon energy dependence of the matrix elements.
All of these new data and their analysis substantially strengthen our
earlier conclusions. (See also ref.~\onlinecite{HELEN}).

In Fig.~12 we show the electronic structure from EDC peaks
(in panel (c)) and Fermi surface crossings (in panel (b)) determined
from data at incident photon energies of 19 and 22 eV for an OD87K
sample\cite{BILAYER}.
The dark lines in the bottom panel of Fig.~12 are a fit of
the intrinsic planar CuO$_2$ electronic
structure, $\epsilon_{\bf k}$, which we call the ``main band''; see
Ref.~\onlinecite{BILAYER,NORM95} for details. The lighter lines
are simply obtained by plotting $\epsilon_{\bf k \pm Q}$,
where ${\bf Q}$ is the superlattice vector, and it is very important
to note that these lines provide an excellent description
of the data points that do not lie on main band. The data strongly
suggest \cite{SLDIFF} that these additional ``umklapp bands''
arise due to diffraction of the outgoing photoelectron through
the BiO superlattice, which leads to ``ghost'' images of the electronic
structure at $\epsilon_{\bf k \pm Q}$.

     From the point of view of the present discussion of the
Fermi crossings, it is very important to establish conclusively
that the crossings U4 and U5 along $(0,0)$ to $(\pi,0)$
shown in the middle panel of Fig.~12 correspond to umklapp
``ghost'' images and {\it not} to the ``main band''.
The case of the Fermi crossing U5, closer to $(0,0)$, is unambiguous
since it is just obtained from following the dispersion of the EDC peaks
which clearly fall on the umklapp band dispersion in Fig.~12(c).

The Fermi crossing U4 requires more care since the umklapp and main
band dispersions are almost degenerate in the vicinity of $(\pi,0)$.
To disentangle these two contributions, we exploit the polarization
selection rules discussed in Section IV.
In Fig.~12(c) we use filled circle symbols to denote data obtained
in an odd geometry, i.e., initial state odd under reflection in the
corresponding mirror plane, and open circles to denote even geometry.
We see from the dispersion plotted in Fig.~12(c) that the main
band signal is seen in the even geometry, since it is a dipole-allowed
transition. (Actually in this polarization both the main and the umklapp
bands should contribute but in the 17 -- 22 eV photon energy range the
main band intensity is much larger than that of the umklapps.)
However in the odd polarization (filled circles along $(0,0)$ to
$(\pi,0)$) the main band is dipole-forbidden and thus the weaker
umklapp band, which does not have any symmetry restrictions here,
dominates in this geometry. From the dispersion, and in particular the
polarization geometry in which it is observed, we clearly see that
the Fermi crossing U4 must be associated with the umklapp
Fermi surface, and {\it not} with the main band.

Fig.~13 shows various cuts at a photon energy of 22 eV for Bi2212-OD87K.
The Fermi crossings are determined using the symmetrization method
and the Fermi surface is found to be a hole barrel
centered at $(\pi,\pi)$.
Notice that in each cut umklapp bands (broken
lines) can be identified. The labels U1 to U3 correspond to particular Fermi
crossings of the umklapp band as shown in Fig. ~12.
In particular along the $(0,0)$ to $(\pi,\pi)$ direction (right-side panels)
the U1 and U2 crossings of the umklapp bands can be
clearly observed.

These umklapp bands are also responsible
for the $\omega=0$ peaks observed in the symmetrized data beyond
${\bf k_F}$ in the other panels.
The new data, obtained at much higher density in the zone, allows us to
directly visualize the main Fermi surface, together with the ghost
Fermi surfaces due to umklapp bands, using the same procedure as in
Fig.~11. The magnitude of logarithmic gradient
$|\nabla_{\bf k}I({\bf k})|/I({\bf k})$ for Bi2212-OPT90K
over part of the Brillouin zone in the Y quadrant (defined
in Fig.~12) is shown in Fig.~14.
From the intensity pattern in this plot, one can clearly see
the main Fermi surface in the middle, 
which is a large hole-like barrel,
and also one of the umklapps (U3 in Fig.~12 notation).
The other umklapps are weaker in intensity, but would be visible in the
figure if a log intensity scale had been used.  It is quite satisfying to
see indications of all the features deduced earlier
(using other methods) in the plots obtained using the straightforward
logarithmic gradient method.

The analysis of the photon energy dependence of the ARPES data in
Bi2212 shows a similar trend to the one observed in Bi2201. As an
example of this, we plot in Fig.~15 the
photon energy dependence of the ratio $I(0.7\pi,0)/I(\pi,0)$
of the energy integrated intensity measured at $(0.7\pi,0)$ and $(\pi,0)$
for OD88K. 
Both these ${\bf k}$ points are inside the occupied part of the
zone, and we would not expect the momentum distribution $n({\bf k})$
to vary significantly from one point to the other.
Thus, following Eq.~(\ref{integratedI}), any significant deviation of
this ratio from unity as a function of incident photon
energy must be attributed to the matrix elements.
While around 20 eV, $I(0.7\pi,0)/I(\pi,0)$ is close
to unity indicating a small ${\bf k}$-dependence of the matrix elements
in this part of the zone, the ratio peaks to about
2.5 at 38 eV, signaling the suppression of intensity around $(\pi,0)$,
similar to what is seen in Bi2201.
Even more interesting is the observation that around 54
eV the $I(0.7\pi,0)/I(\pi,0)$ becomes much {\it smaller} than one.
Fig.~15 
illustrates once more how dangerous it would be to infer a Fermi
crossing from intensity variations alone.

We have further observed \cite{HELEN} that at photon energies close to
30 eV, where the main band is strongly suppressed around $(\pi,0)$,
the superlattice contributions are in fact strongly enhanced in this
region. 
The reason is that these superlattice intensities originate from regions
of reciprocal space where the matrix elements are less suppressed.

To summarize, it is very important that
the superlattice contributions be differentiated from the main band
using polarization selection rules, and all Fermi crossings carefully
checked by a combination of symmetrization analysis together with careful
studies of the integrated intensities as shown above.

\section{Conclusions}

In this paper we have carefully enunciated the criteria to
be used in determining the Fermi surface from ARPES data.
We have illustrated these ideas using data on two high $T_c$ copper oxide
based materials, Bi2201 and Bi2212. However we believe that
these methods should prove to be useful for a large class of
quasi 2D materials.

The high $T_c$ materials are hard to analyze because of the
absence of sharp quasiparticle peaks in their normal state,
and anomalously weak dispersion in parts of the zone.
Nevertheless, the symmetrization method discussed in this
paper is able to deal with both these issues. It effectively
removes the Fermi function from the EDC and determines
${\bf k}_F$ as that point in ${\bf k}$-space at which the
spectral function peaks at the chemical potential.

It is very useful to supplement this analysis with
studies of the momentum distribution, however one has
to be very careful about matrix element effects.
We show that by analyzing data at different incident photon energies,
one can unambiguously distinguish between loss of integrated intensity
arising from matrix element variations from that due to genuine structure
in
$n({\bf k})$. In this connection we have also shown the
usefulness of studying the gradient of the logarithm of
the integrated intensity.

We emphasize that not recognizing the role of matrix elements can lead to
paradoxical conclusions like changes in Fermi surface topology with photon
energy, which of course makes no sense. Some authors have ascribed the
differences between the 22 eV and 34 eV photon energy data to
additional states around $(\pi,0)$ \cite{DESSAU99,SHEN99}.
We find that there is no necessity to invoke such states.

Bi2212 has an additional complication arising from the effect
of the BiO superlattice modulation on the ARPES data. However,
as we argue above, the Fermi surface crossings arising from
``ghost'' images (superlattice umklapp bands) can be
clearly differentiated from the intrinsic planar CuO$_2$
Fermi surface by exploiting polarization selection rules.

In conclusion, all our data on Bi2201 and Bi2212
when analyzed using the methods
described above indicate a single large Fermi surface centered around
$(\pi,\pi)$ \textit{independent of the photon energy}.

\acknowledgments

We thank Jim Allen for discussions which led us to sharpen some
of the arguments presented in the Appendix.
This work was supported by the National Science Foundation DMR 9624048,
and DMR 91-20000 through the Science and Technology Center for
Superconductivity, the U. S. Dept. of Energy, Basic Energy Sciences,
under contract W-31-109-ENG-38, the CREST of JST,
the Ministry of Education, Science, and Culture of Japan, and
the Swiss National Science Foundation. MR is supported in
part by the Indian DST through the Swarnajayanti scheme.

\appendix
\section{Symmetrization and Resolution}

In this Appendix we discuss in detail the effect of
experimental resolution on the symmetrization method.
For simplicity, we discuss the
elimination of the Fermi function from the EDC {\em at} ${\bf k}_F$,
which requires an assumption of particle-hole symmetry. As discussed
in Section V, no such assumption was required for the determination
of ${\bf k}_F$ via symmetrization.

Symmetrization was first introduced by us in
Ref.~\onlinecite{NORMAN98} and used extensively
for studying the self energy in Ref.~\onlinecite{PHENO}.
The main result 
\begin{equation}
I_{\rm sym}(\textbf{k}_F,\omega)
= I(\textbf{k}_F,\omega) + I(\textbf{k}_F,-\omega)
= I_0 A(\textbf{k}_F,\omega).
\label{symm0x}
\end{equation}
follows immediately from Eq.~(\ref{intensity}) by using the identity
\begin{equation}
f(-\omega) = 1 - f(\omega)
\label{fermi}
\end{equation}
obeyed by the Fermi function,
together with the assumption of particle-hole symmetry at low energies
($\omega$ less than few times the temperature):
\begin{equation}
A({\bf k}_F,\omega) = A({\bf k}_F,-\omega).
\label{ph1}
\end{equation}

Let us now see how symmetrization works in the presence of finite
energy and momentum resolutions.
For clarity of presentation, we discuss these one at a time,
although both can be trivially treated together.
With a finite energy resolution, Eq.~(\ref{intensity}) is generalized to
\begin{equation}
I({\bf k},\omega) = I_0 \int_{-\infty}^{+\infty}d\omega^\prime
R(\omega - \omega^\prime)f(\omega^\prime)A({\bf k},\omega^\prime),
\end{equation}
where $R$ is typically taken to be a Gaussian.
Using $R(\omega - \omega^\prime) = R(\omega^\prime - \omega)$
and Eqs.~(\ref{fermi}) and (\ref{ph1}), we can easily see that
\begin{eqnarray}
I_{\rm sym}(\textbf{k}_F,\omega)
= I(\textbf{k}_F,\omega) + I(\textbf{k}_F,-\omega) \nonumber \\
= I_0 \int_{-\infty}^{+\infty}d\omega^\prime
R(\omega - \omega^\prime) A(\textbf{k}_F,\omega^\prime).
\label{symm1}
\end{eqnarray}
Thus symmetrization succeeds in removing the effect of the
Fermi function from inside the convolution integral.

Next consider the effect of a small, but finite, ${\bf k}$-window.
In its presence, Eq.~(\ref{intensity}) is replaced by
\begin{equation}
I({\bf k},\omega) = I_0 f(\omega) \sum{ }^\prime A({\bf k}^\prime,\omega).
\label{kwindow}
\end{equation}
where $\sum^\prime$ is shorthand for summation over ${\bf k}^\prime$
within the window centered about ${\bf k}$. We ignore the
${\bf k}$-variation of the prefactor $I_0$ within this small window
which allows us to pull it out of the sum. Next we need to extend our
particle-hole symmetry assumption to ${\bf k}$'s slightly away from ${\bf
k}_F$.
We require
\begin{equation}
A(\epsilon_{\bf k},\omega) = A(-\epsilon_{\bf k},-\omega)
\label{ph2}
\end{equation}
valid for $|\omega|$ and $|\epsilon_{\bf k}|$ both less than few tens of
meV.
Note that we have rewritten the first argument of the spectral function as
$\epsilon_{\bf k}$, which can be linearized in the vicinity
of ${\bf k}_F$ as $\epsilon_{\bf k} \simeq {\bf v}_F.\delta{\bf k}$
where $\delta{\bf k} = ({\bf k} - {\bf k}_F)$.
The symmetrized intensity is thus given by
$I_{\rm sym}(\textbf{k}_F,\omega) =
I_0 f(\omega) \sum^\prime A({\bf v}_F.\delta{\bf k},\omega)
+ I_0 f(-\omega) \sum^\prime A({\bf v}_F.\delta{\bf k},-\omega)$.
We assume a {\it symmetric} window, so that if ${\bf k}_F + \delta{\bf k}$
is within the window, then so is ${\bf k}_F - \delta{\bf k}$.
Thus we can rewrite the second term as
$\sum^\prime A({\bf v}_F.\delta{\bf k},-\omega)
= \sum^\prime A(-{\bf v}_F.\delta{\bf k},-\omega)$,
which on using Eq.~(\ref{ph2}) is given by
$\sum^\prime A({\bf v}_F.\delta{\bf k},\omega)$.
Finally, using Eq.~(\ref{fermi}), we get
\begin{equation}
I_{\rm sym}(\textbf{k}_F,\omega) =
I_0 \sum{ }^\prime A({\bf k}^\prime,\omega).
\label{symm2}
\end{equation}

Combining the arguments that led to Eqs.~(\ref{symm1}) and (\ref{symm2}),
we see that the symmetrization procedure works in the presence of both
energy and momentum resolution.
In our previous work \cite{NORMAN98,PHENO} we had used this
procedure to analyze data at ${\bf k}_F$ mostly
in the pseudogap and superconducting states.

\begin{figure}
\caption{
Rocking curve of a (0,0,10) reflection on a
Bi2212 sample showing the large structural
coherence length.
}
\label{fig1}
\end{figure}

\begin{figure}
\caption{
Energy distribution curves (EDCs) obtained at $h\nu=$22 eV
for a Bi2201-OD4K sample,
showing the symmetry about
the $(0,0) \to (2\pi,0)$ (a) and the $(\pi,-\pi)
\to (\pi,\pi)$ (b) directions.}
\label{fig2}
\end{figure}

\begin{figure}
\caption{
(a) Electronic structure of the valence band of Bi2212 obtained
by taking the second derivative of EDCs such as those shown in (b)
at $h\nu=$22 eV.
}
\label{fig3}
\end{figure}

\begin{figure}
\caption{
(a) Arrangement of the photon beam and detector in
order to make use of the photoemission selection rules.
(b) Parity of the Cu$d_{x^{2}-y^{2}}$ orbitals hybridized with the
O$2p$ orbitals. (c) EDCs showing the parity of the
orbitals in (b) obtained at $h\nu=$22 eV.}
\label{fig4}
\end{figure}

\begin{figure}
\caption{
Determination of the Fermi crossing along
$(0,0) \to (\pi,\pi)$ in Bi2212-OD88K obtained at $h\nu=$22 eV, T=180~K:
(a) EDCs with the state at the Fermi
momentum shown in bold; (b) Symmetrization of the EDCs in (a); (c)
Dispersion obtained from the data in (a).
}
\label{fig5}
\end{figure}

\begin{figure}
\caption{
EDCs from Bi2201-OD23K obtained at $h\nu=$22 eV, T=25~K along cuts
perpendicular to the $(0,0) \to (\pi,0)$ direction, with $k_{x}$
indicated 
in each panel.}
\label{fig6}
\end{figure}

\begin{figure}
\caption{
(a,b,c) Symmetrization of selected data in Fig.~6.
(d) EDCs along the $(0,0) \to (2\pi,0)$ direction; (e)
symmetrization of the EDCs in (d), showing no
Fermi surface crossing.}
\label{fig7}
\end{figure}

\begin{figure}
\caption{
Determination of the Fermi crossing in Bi2201-OD23K
at $h\nu=$34 eV, T=25~K. (a) EDCs along the $(0,0) \to (\pi,0) \to
(\pi,\pi)$ directions. Symmetrized data (b) along $(0,0) \to
(\pi,0)$
showing no Fermi crossing and (c) along $(\pi,0) \to (\pi,\pi)$ showing a
clear Fermi crossing.
}
\label{fig8}
\end{figure}

\begin{figure}
\caption{
Bi2201-OD23K, (a) Integrated intensity at $h\nu=$22 and 34 eV
along the $(0,0) \to (\pi,0) \to (\pi,\pi)$ directions. (b)
Comparison of the ARPES lineshape measured at 22 (dashed lines) and 34 eV
(solid lines) at three different ${\bf k}$ points.
}
\label{fig9}
\end{figure}

\begin{figure}
\caption{
(a) Model calculation of the dispersive band along
the $(0,0) \to (\pi,0)$ direction, assuming no (dashed lines) and
strong (solid lines) k-dependence of the matrix elements. The
variation of the matrix element observed in Fig.~9 at 34 eV is
simulated by a $\sin^{4}(0.6k_{x})$ function. (b) Integrated
intensity over a narrow (-50,+50 meV) and wide (-350, +50 meV)
energy range using the model
calulation shown in (a). (c) Integrated intensity over the same
integration ranges as in (b) obtained
experimentally on Bi2201-OD23K at $h\nu=$34 eV.
}
\label{fig10}
\end{figure}

\begin{figure}
\caption{
Bi2201-OD0K: (a and c) Integrated intensity (over -310 to +90 meV)
$I({\bf k})$
measured at $h\nu=$22 and 28 eV, T=16~K around the $(\pi,0)$ point. Notice
that the intensity maximum depends strongly upon the photon energy
$h\nu$.
(b and d) Corresponding gradient of the logarithm,
$|\nabla_{\bf k}I({\bf k})|/I({\bf k})$, the maxima (white) which
correspond to Fermi crossings and clearly show that, \textit{independent
of 
the photon energy}, the Fermi surface consists of a hole barrel centered
around $(\pi,\pi)$.
}
\label{fig11}
\end{figure}

\begin{figure}
\caption{
(a) In the presence of the superstructure in the BiO layers, the
outgoing electrons can be diffracted, thus giving rise to additional
umklapp bands as shown in (b and c). (b) Main (thick) and
umklapp (thin) Fermi surfaces. Selected Fermi crossings of the umklapp
bands relevant for Figs.~13 and 14 are labeled from U1 to U5.
(c) Dispersions obtained from earlier measurements on
Bi2212-OD87K\protect\cite{BILAYER}.
Filled circles denote data obtained in an odd polarization, i.e.,
initial state odd under reflection in the corresponding mirror plane.
Open circles denote even polarization, and open triangles correspond
to a mixed polarization.
}
\label{fig12}
\end{figure}

\begin{figure}
\caption{
Bi2212-OD87K. The top panels show the EDC cuts taken at $h\nu=$22
eV, T=100~K together with the
polarization geometry used. Various EDCs are shown in the middle panels,
together with the corresponding symmetrized data in the lower panels. The
curves corresponding to the Fermi
crossing of the main and umklapp bands are shown with thick and
broken lines, respectively.
The labels U1 to U3 correspond to
particular Fermi crossings of the umklapp
band as shown in Fig. ~12.
}
\label{fig13}
\end{figure}

\begin{figure}
\caption{
Gradient of the logarithm of the integrated intensity (over -320 to +80
meV), 
$|\nabla_{\bf k}I({\bf k})|/I({\bf k})$, around
$(\pi,0)$ for Bi2212-OPT90K taken at $h\nu$=22 eV.
Note the large hole-like Fermi surface corresponding to the
main band. In addition, the umklapp band (U3 in Fig.~12 notation) 
can be seen as well. The other umklapps are weaker in intensity, 
but would be visible if a log intensity scale had been used instead.
}
\label{fig14}
\end{figure}

\begin{figure}
\caption{
Photon energy dependence of the integrated intensity ratio (over -600 to
+200
meV) $I(0.7\pi,0)/I(\pi,0)$ of Bi2212-OD88K.
The ratio is near unity at 20 eV, increases
at 30 eV to two and larger, and decreases at 54 eV to less than one-half.
This figure illustrates how dangerous it would be to
infer a Fermi crossing from intensity variations only.
}
\label{fig15}
\end{figure}

\begin{figure}
\caption{
Plot of symmetrized intensities obtained from simulation for five ${\bf k}$
points ($\phi=2.5^\circ - 6.5^\circ$) for a $\theta=14^\circ$ cut in
${\bf k}$-space ($\Gamma M$ geometry).  The curve closest to $k_F$
($\equiv 4.6^\circ$) is shown in bold.  The dispersion was chosen from the
tight-binding fit of Ref.~\protect\onlinecite{NORM95}.
A (constant) linewidth broadening of 50 meV was used, with a
Gaussian energy resolution of $\sigma = 15 meV$, and a $1^\circ$ radius
${\bf k}$-window. A Fermi function with $T=14$ K was used, and
matrix elements were ignored. Using these parameters, the five EDCs were
generated and then plotted after symmetrization.
}
\label{fig16}
\end{figure}


\begin{references}
\bibitem{INTENSITY}
We note that the intensity may not be the same on either side of
the $(\pi,0)$ point because of either matrix element effects or
the direction of incident light.
\bibitem{EXPT}
For this discussion it is perhaps simpler to imagine changing the photon
polarization, keeping the detector fixed.
In the actual experiment, however,
it is the photon polarization which is fixed and
the detector is moved when checking dipole selection rules.
\bibitem{SLDIFF}
There are two different interpretations of the umklapp ``bands'':
(a) a final state effect involving diffraction of the outgoing
photoelectron off the BiO superlattice,
or (b) an initial state effect caused by the
BiO superlattice perturbing the CuO$_2$ electronic structure.
There are several pieces of evidence in favor of interpretation (a).
First, polarization selection rules are obeyed for the $(0,0)-(\pi,0)$
mirror plane, and this would not be a mirror plane if the initial state
electronic structure was affected by the superlattice.
Second, the intensities of the main band compared with the umklapp bands
in the vicinity of $(\pi,0)$
argue strongly in favor of interpretation (a);
see ref.~\onlinecite{HELEN}.
\end{references}
\end{document}